\documentclass[runningheads]{llncs}
\usepackage{graphicx}
\usepackage{amsmath}

\usepackage{enumitem}
\usepackage{multirow}
\usepackage{makecell}
\usepackage{colortbl} 
\usepackage[table]{xcolor}
\definecolor{lightgray}{rgb}{0.9, 0.9, 0.9}

\begin{document}
\title{CT-based brain ventricle segmentation via diffusion Schrödinger Bridge without target domain ground truths}
%
%
\author{Reihaneh Teimouri\inst{1} \and
Marta Kersten-Oertel\inst{1}\and
Yiming Xiao\inst{1}}
%

\authorrunning{Teimouri et al.}
%
\institute{Department of Computer Science and Software Engineering, Concordia University, Montreal, Canada\\
\email{reihaneh.teimouri@mail.concordia.ca}\\
\email{\{marta.kersten, yiming.xiao\}@concordia.ca}\\
}
\maketitle              
\begin{abstract}
Efficient and accurate brain ventricle segmentation from clinical CT scans is critical for emergency surgeries like ventriculostomy. With the challenges in poor soft tissue contrast and a scarcity of well-annotated databases for clinical brain CTs, we introduce a novel uncertainty-aware ventricle segmentation technique without the need of CT segmentation ground truths by leveraging diffusion-model-based domain adaptation. Specifically, our method employs the diffusion Schrödinger Bridge and an attention recurrent residual U-Net to capitalize on unpaired CT and MRI scans to derive automatic CT segmentation from those of the MRIs, which are more accessible. Importantly, we propose an end-to-end, joint training framework of image translation and segmentation tasks, and demonstrate its benefit over training individual tasks separately. By comparing the proposed method against similar setups using two different GAN models for domain adaptation (CycleGAN and CUT), we also reveal the advantage of diffusion models towards improved segmentation and image translation quality. With a Dice score of 0.78±0.27, our proposed method outperformed the compared methods, including SynSeg-Net, while providing intuitive uncertainty measures to further facilitate quality control of the automatic segmentation outcomes. The code is available at: \url{https://github.com/HealthX-Lab/DiffusionSynCTSeg}.

\end{abstract}

\section{Introduction}
\label{sec:Introduction}
Ventriculostomy plays a vital role in managing conditions like hydrocephalus, traumatic brain injury, and brain tumors by draining excessive cerebrospinal fluid (CSF) through a catheter inserted into the brain's ventricles. The procedure, often performed in emergent settings, requires good localization of the ventricular anatomy. Most commonly the freehand technique, where external anatomical landmarks are used to determine the entry point, is used.  However, this technique is less precise due to sometimes difficult-to-discern landmarks, and is associated with over 30\% placement error \cite{external_ventricular_drain}. Pre-procedure CT scans can help guide the procedure, and automated ventricle segmentation can significantly improve treatment planning and catheter placement precision \cite{ventricle-segmentation}. Automatic brain ventricle segmentation in MRI has progressed from label fusion to advanced deep learning (DL), but CT faces challenges, due to low soft tissue contrast and limited quality, especially in clinical CT scans where lower radiation dosage is preferred. This is further complicated by the lack of well-annotated public CT datasets for relevant DL model development and validation.  

\begin{figure}[ht]
\includegraphics[width=\textwidth]{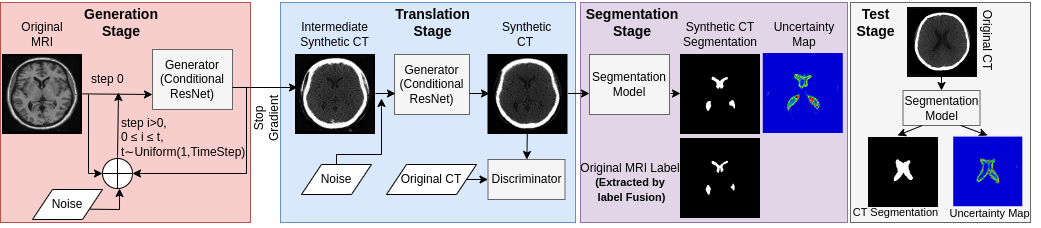}
\caption[An overview of the UNSB Model]{Overview of the proposed uncertainty-aware brain ventricle segmentation technique, based on Unpaired Neural Schrödinger Bridge (UNSB) and joint image translation and segmentation training. Note that in the test stage, only the segmentation module is used to obtain segmentation in real CT scans.} \label{fig:diffmodel}
\end{figure}

\begin{figure}[ht]
\includegraphics[width=\textwidth]{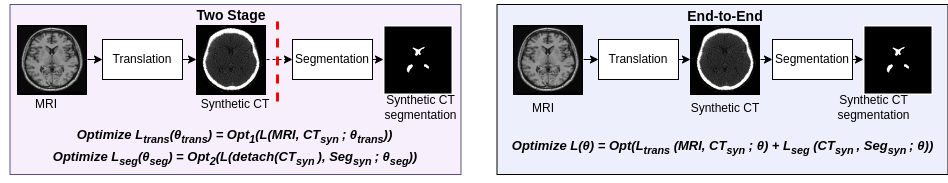}
\caption[E2E]{An overview of two different training approaches for ventricle segmentation. \emph{Left:} Two-stage approach with image translation and segmentation modules trained separately. \emph{Right:} End-to-end approach with a unified loss function for both tasks.} \label{fig:E2E}
\end{figure}

To address the aforementioned challenges in CT-based brain ventricle segmentation, we developed an end-to-end DL technique that employs a diffusion model to allow MRI-to-CT domain adaptation without target domain ground truths (see Fig. \ref{fig:diffmodel}). Specifically, we leverage more accessible MRI ground truth labels to learn synthesized segmentation in unpaired CT scans by performing joint learning of image translation and segmentation tasks. Unpaired image contrast/modality translation faces challenges in unmatched features. Considering the recent success of diffusion models \cite{diffusion} and training stability issues of the commonly used CycleGAN \cite{CycleGAN} framework, we employ the Unpaired Neural Schrödinger Bridge (UNSB) \cite{unpairedDiff} approach, which allows for high-fidelity unpaired MRI vs. CT translation. For the segmentation component, we utilized an attention recurrent residual U-Net (R2AUNet)\cite{R2AU-Net} and Monte Carlo (MC) dropout\cite{Monte-Carlo} to allow uncertainty assessment for intuitive quality control of the results. Our work has three main contributions. \textbf{First}, we proposed joint learning of diffusion-based unpaired MRI-to-CT translation and segmentation for the first time, and demonstrated its benefit over two-stage training approaches (see Fig.~\ref{fig:E2E}). \textbf{Second}, we performed a thorough comparison of our approach against similar setups with two state-of-the-art Generative Adversarial Network (GAN)-based techniques, including CycleGAN\cite{CycleGAN} and Contrastive Learning for Unpaired Image-to-Image Translation (CUT) \cite{CUT_UDA} to show the benefits of the diffusion Schrödinger Bridge. \textbf{Lastly}, we incorporated visual uncertainty estimation metrics in the image segmentation module to enhance the interpretability of the automatic segmentation quality.

\section{Related Works}
\label{sec:Related-Works}
Volumetric quantification of the brain ventricles is a crucial biomarker for brain development, aging, and cognitive decline. Thus, ventricle segmentation in MRI has attracted significant interest in the neuroscience community, with methods ranging from classic and robust label-fusion \cite{labelfusion} to efficient DL-based techniques \cite{DA_ventricle_seg,DLventriculostomy}. CT-based ventricle segmentation remains less developed, often requiring significant in-house data and resources \cite{DLventriculostomy}. To leverage more accessible resources from MRI segmentation, different strategies, such as GANs for domain adaptation have been used. However, these typically require labeled data and/or paired samples across both domains \cite{DomainAdaptationSurvey}. Mitigating the need for paired image contrasts/modalities, CycleGANs \cite{CycleGAN} that employ a cycle consistency loss have been used in SynSeg-Net \cite{Synseg} to obtain synthetic segmentation through joint image translation and segmentation learning without target domain ground truths. Due to the difficulty in CycleGANs training, other approaches have also been proposed. For example, Contrastive Learning for Unpaired Image-to-Image Translation (CUT) \cite{CUT_UDA} is a GAN method that uses patch-based contrastive learning to enforce content correspondence across domains. Originally proposed for natural images, it provides an elegant framework to allow unpaired image translation with improved feature preservation. In medical images, Choi et al. utilized the CUT model for vestibular schwannoma and cochlea segmentation, achieving an accuracy of 0.82 with unpaired data \cite{choi2021using}. Wang et al. \cite{jiangtao2021mri} performed MRI-to-CT synthesis with the CUT model, achieving a 0.91 structural similarity index measure (SSIM) with paired data. Diffusion models\cite{diffusion_model} have also been adopted in medical imaging applications. Ozbey et al. \cite{ozbey2023unsupervised} presented SynDiff for image contrast and modality translation in MRI and CT. Also, Graf et al. \cite{graf2023denoising} applied denoising diffusion models for spinal MRI-to-CT translation, notably improving segmentation accuracy for spinal structures. Most recently, to overcome the limitation in unpaired high-resolution image translation with diffusion models due to the simple Gaussian prior assumption, Kim et al. \cite{unpairedDiff} proposed the Unpaired Neural Schrödinger Bridge (UNSB), which learns a stochastic differential equation to bridge two arbitrary distributions as a sequence of adversarial learning problems. This technique has not been adopted for medical imaging tasks. In our work, we employed the UNSB in a joint image translation and segmentation learning framework for ventricle segmentation for the first time.

\section{Methods and Materials}
\subsection{Dataset and preprocessing}
\label{sec:Data}
We curated T1w MRI and CT scans from diverse sources, including the CERMEP-iDB-MRXFDG (iDB, 27 sets of paired MRIs and CTs) \cite{iDB}, OASIS3 (200 CTs)\cite{OASIS3}, NeuroMorphometrics\footnote[1]{http://www.neuromorphometrics.com} (68 MRIs from OASIS \cite{OASIS1} and ADNI\footnote[3]{adni.loni.usc.edu}), IXI \footnote[2]{https://brain-development.org/ixi-dataset/}(63 MRIs), and an in-house dataset (67 CTs). As the only dataset with paired and co-registered MRI-CT scans, iDB data were reserved as the test set, and the remaining data (267 CT and 131 MRI volumes, unpaired) were employed for model development. Consistent with \cite{70Elderly}, our focus was on subjects over 70 years of age. All MRI scans underwent N4 bias correction \cite{N4ITK}, while we used a window of [-10,500] HU for the CT scans. All images were linearly aligned in the ICBM152 space\cite{icbm2} with 1 mm isotropic resolution. For MRI ventricle segmentation, for ground truths we used expert annotations from the NeuroMorphometrics dataset. For the iDB and IXI datasets, we generated silver ground truths using multi-atlas label fusion and majority voting, utilizing the entire NeuroMorphometrics dataset as our atlas library \cite{Parkinson_seg}. Quality control was conducted by the first author. \textit{Note that label-fusion for CT would be challenging due to poor soft tissue contrast}. We used 2D axial slices of 256$\times$256 pixels that were normalized to [-1, 1] for model training and validation.

\subsection{Network architecture and loss function}
\label{sec:Networks}

We employed the UNSB model alongside an R2AUNet segmentation module to tackle high-resolution, unpaired image translation challenges (see Fig. \ref{fig:diffmodel}). The UNSB model stands out by utilizing stochastic optimization and a diffusion-based generative process, guided by a sophisticated loss framework comprising adversarial ($L_{\text{Adv}}$), Schrödinger Bridge ($L_{\text{SB}}$), and regularization ($L_{\text{Reg}}$) losses. This setup facilitates a nuanced transition between source and target domains by creating intermediate representations that gradually adopt the target style. Central to UNSB's operation is the Neural Schrödinger Bridge principle, which strategically minimizes the energy gap between source and target image distributions. It achieves this by mapping out an optimal transformation pathway that maintains the original images' semantic content while seamlessly integrating the stylistic elements of the target domain, effectively bridging the gap between domains without the need for paired examples. Specifically, the UNSB\cite{unpairedDiff} uses time-conditional neural networks, which discretize the transformation process into manageable segments ($\{t_i\}_{i=0}^{N}$, with $N=5$), ensuring a precise and controlled approach to image synthesis and translation. The loss function for UNSB is,
\begin{equation}
L_{\text{UNSB}}(\phi, t_i) = L_{\text{Adv}}(\phi, t_i) + \lambda_{\text{SB}, t_i} L_{\text{SB}}(\phi, t_i) + \lambda_{\text{Reg}, t_i} L_{\text{Reg}}(\phi, t_i)
\end{equation}
where, $t_i$ represents the time step, and $\phi$ denotes the generator model. In this formula, \(\lambda_{\text{SB}}\) and \(\lambda_{\text{Reg}}\) are both equal to 1. In the generator, we used a conditional ResNet model with a 9 blocks architecture\cite{unpairedDiff}. 

For segmentation, we utilized an Attention Recurrent Residual Convolutional U-Net (R2AUNet)\cite{R2AU-Net}, and compared it against a conventional U-Net\cite{unet}. We addressed dataset imbalances and focused on precise ventricle detection by employing a weighted cross-entropy loss, assigning a weight of 30 to the ventricle class due to its importance and the abundance of background pixels. R2AUNet improves generalization and mitigates the vanishing gradient problem through recurrent and residual connections. We explored two CT image segmentation paradigms: end-to-end and two-stage, as depicted in Fig. \ref{fig:E2E}, and marked as ``E2E'' and ``2Stage'' in Table~\ref{tabCUT}, respectively. The end-to-end method combines translation and segmentation, as detailed in Equation \ref{eq:UNSBsegloss}. The two-stage approach uses distinct optimizers to minimize errors for individual tasks. Upon completing the training for the image translation module, this paradigm employs all resulting synthetic images to train the segmentation network.
\begin{equation}
L_{\text{E2E UNSB+Segmentation}} =  L_{\text{UNSB}}  + L_{\text{segmentation}}
\label{eq:UNSBsegloss}
\end{equation}
We incorporated uncertainty estimation using Monte Carlo Dropout\cite{Monte-Carlo} in the segmentation model for effective quality control. The dropout layer was added before the last convolutional layer of the decoder in the segmentation model (ResNet for SynSeg-Net and R2AUNet/UNet for other models, see Section 3.3). During testing, we run 10 forward passes (dropout rate = 0.5) to create distributions of outcomes for each CT scan. Computing the mean values and variances across these outcomes generates the final predictions and associated uncertainty maps. These maps highlight areas needing closer scrutiny by clinicians. This technique quantifies epistemic uncertainty\cite{amini2020deep}, offering valuable insights into the model's confidence levels in predictions and aiding clinicians in identifying ambiguous regions for further examination.

\subsection{Experimental setup and method comparisons}
\label{sec:setup}

To fully validate the proposed diffusion-based segmentation method with end-to-end training (E2E-UNSB+R2AUNet), we conducted a comprehensive comparison with other models. Specifically, we performed experiments to evaluate different aspects of our model design by establishing a series of other DL models for the same task (see Table 1). \textbf{First}, to determine the benefit of UNSB for image translation in joint image-translation and segmentation training, we built a similar model based on CUT (E2E-CUT+R2AUNet) and CycleGAN (SynSeg-Net\cite{Synseg}). \textbf{Second}, to validate the advantages of joint image translation and segmentation training, we compared the end-to-end models with UNSB and CUT against two-stage models, namely 2Stage-UNSB+R2AUNet and 2Stage-CUT+R2AUNet, where image-translation and segmentation modules were trained separately. \textbf{Third}, to determine the effectiveness of the attention recurrent residual U-Net as the segmentation module, models (E2E-UNSB+UNet, 2Stage-UNSB+UNet, and 2Stage-CUT+UNet) based on plain 2D UNets\cite{unet} were tested. \textbf{Finally}, as an alternative strategy, CT segmentation based on MRI ground truths can be done by first passing the input CT scan through a CT-to-MRI translation network, and then passing the synthetic MRI to an MRI segmentation network (trained using MRI ground truth labels). Thus, we trained two additional DL models, including 2Stage-CT-to-MRI-UNSB+R2AUNet and 2Stage-CT-to-MRI-CUT+R2AUNet. Our models were built using PyTorch 1.13.1 and trained using the Adam optimizer with a learning rate of 0.0002. We applied a batch size of one for UNSB-based and CUT-based models and two for SynSeg-Net, all running on a Tesla V100-SXM2 GPU with 32GB of memory. Here, for the reported optimal results, E2E-UNSB+R2AUNet, E2E-CUT+R2AUNet, and SynSeg-Net  were trained for 48 epochs, 45 epochs, and 100 epochs, respectively. 

\begin{table}[t]
\centering
\scriptsize
\rowcolors{2}{gray!15}{gray!0}
\renewcommand{\arraystretch}{1.1}
\begin{tabular}{|c|c|c|c|c|c|}
\hline
Model & \makecell{DSC } & \makecell{IoU }  &\makecell{MSE} & \makecell{SSIM} & \makecell{DSC 3D}\\

        \hline
\textbf{E2E-UNSB+R2AUNet} & \textbf{0.78$\pm$0.27} & \textbf{0.70$\pm$0.29} & 0.11$\pm$0.02 & 0.71$\pm$0.02 & \textbf{0.73$\pm$0.08}\\

2Stage-UNSB+R2AUNet  & 0.63$\pm$0.32 &  0.55$\pm$0.36 & 0.10$\pm$0.01 & 0.72$\pm$0.02 & 0.47$\pm$0.10  \\

E2E-UNSB+UNet  &  0.73$\pm$0.30 & 0.66$\pm$0.32 & 0.10$\pm$0.01 & 0.68$\pm$0.01 & 0.67$\pm$0.09 \\

2Stage-UNSB+UNet  &  0.72$\pm$0.30 & 0.64$\pm$0.30  & 0.09$\pm$0.01 & 0.69$\pm$0.02 & 0.65$\pm$0.12 \\
                    
{\makecell{ 2Stage-CT-to-MRI UNSB+R2AUNet}} &  0.63$\pm$0.35 & 0.55$\pm$0.37 & 0.05$\pm$0.01 & 0.55$\pm$0.05 & 0.51$\pm$0.11 \\
                               


\noalign{\hrule height 1.3pt}
E2E-CUT+R2AUNet  & 0.76$\pm$0.26 &  0.68$\pm$0.29 & 0.12$\pm$0.02 & 0.67$\pm$0.02 & 0.69$\pm$0.07 \\

2Stage-CUT+R2AUNet  & 0.73$\pm$0.29  & 0.66$\pm$0.32 & 0.11$\pm$0.02 & 0.68$\pm$0.01 & 0.67$\pm$0.11 \\
                        
2Stage-CUT+UNet  & 0.72$\pm$0.28 &  0.63$\pm$0.32  & 0.42$\pm$0.05 & 0.68$\pm$0.02 & 0.65$\pm$0.09 \\

{\makecell{ 2Stage-CT-to-MRI CUT+R2AUNet}}  & 0.37$\pm$0.48 &  0.37$\pm$0.48  & 0.04$\pm$0.01 & 0.63$\pm$0.04 & 0.18$\pm$0.04 \\                         

\noalign{\hrule height 1.3pt}
SynSeg-Net & 0.63$\pm$0.33 & 0.55$\pm$0.35  & 0.28$\pm$0.02 & 0.56$\pm$0.03 & 0.56$\pm$0.09  \\

\hline
\end{tabular}

\vspace{7pt}
\caption[Comparative analysis of the performance of the proposed UNSB-based and CUT-based model]{Image segmentation and synthesis quality evaluations (mean$\pm$std) for different DL models. Note that MSE and SSIM for CT-to-MRI translation were computed for MRIs instead of CTs. Best segmentation results are in bold fonts.}\label{tabCUT}
\end{table}

\subsection{Evaluation metrics}
\label{sec:metrics}
To evaluate the DL models, we used the Dice similarity coefficient (DSC) and Intersection over Union (IoU) for segmentation accuracy, and mean squared error (MSE) and structural similarity index measure (SSIM) for the quality of synthetic scans depending on the algorithm setup. Specifically, for models with MRI-to-CT translation, MSE and SSIM were computed between real and synthesized CTs while for the two models with CT-to-MRI translation, these metrics were computed for real and synthetic MRIs. As the proposed and baseline methods operate in 2D, we assessed the segmentation accuracy in a slice-by-slice manner, and in addition, we also provided Dice scores based on 3D volumetric labels by concatenating all segmented slices together for evaluation. Finally, two-sided, paired-sample t-tests were performed for the results to confirm the performance of the proposed method against other baselines. 

\section{Results}
\label{sec:Results}

\begin{figure}[ht]
\includegraphics[width=\textwidth]{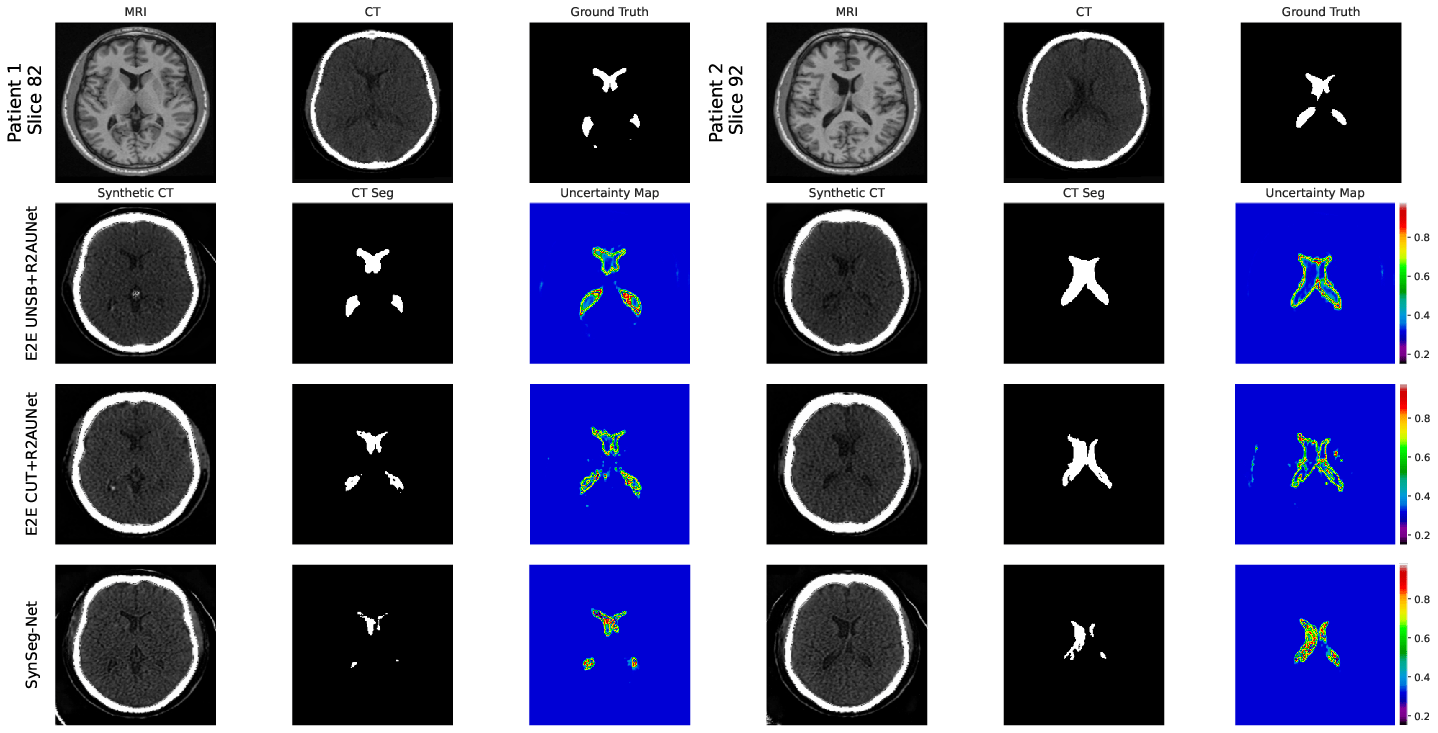}
\caption[2D slice MRI-to-CT translation in all models ]{Qualitative demonstration of ventricle segmentation and uncertainty measures for two subjects with two different slices using different end-to-end methods.} \label{compare_all}
\end{figure}

Quantitative segmentation and image translation evaluations across models are shown in Table~\ref{tabCUT}, with qualitative comparisons in Fig.~\ref{compare_all}.
Overall, our proposed method (E2E-UNSB+R2AUNet), which achieved a Dice score of 0.78±0.27, an IoU of 0.70±0.29, and a 3D Dice score of 0.73±0.08, significantly outperformed the other segmentation models(p$<$0.05). The UNSB model demonstrated a clear advantage over the end-to-end approaches (E2E-UNSB+R2AUNet, E2E-CUT+R2AUNet, and SynSeg-Net). When comparing the ``end-to-end'' and ``two-stage'' approaches with ``UNSB vs. CUT" and ``R2AUNet vs. UNet" settings, we observed better segmentation accuracy (p$<$0.05) for the end-to-end models while the two-stage models have a slight edge in terms of the quality of MRI-to-CT translation. In the segmentation module, the R2AUNet was compared against the vanilla UNet in both ``end-to-end'' and ``two-stage'' approaches, and the first consistently provided superior outcomes. Noticeably, between E2E-UNSB+R2AUNet and E2E-UNSB+UNet, the 2D Dice score, IoU and 3D Dice score were increased by 5\%, 4\%, and 6\%, respectively by the R2AUNet. Finally, although performing CT-to-MRI translation is an alternative method to achieve ventricle segmentation, it significantly reduced the segmentation performance, with heavier impacts on models based on CUT. To validate segmentation uncertainty, we correlated DSC with uncertainties across all test slices. E2E-UNSB+R2AUNet showed a stronger correlation (-0.86) compared to E2E-CUT+R2AUNet and SynSeg-Net (-0.81 and -0.41, respectively).

\section{Discussion}
\label{sec:Discussion}

Our work was inspired by the SynSeg-Net\cite{Synseg}, which conducts joint training of CycleGAN-based unpaired image translation and segmentation. To further enhance the results and accommodate the challenges in clinical head CT scans, we adopted the UNSB framework\cite{unpairedDiff} for medical image synthesis for the first time while employing an attention recurrent residual U-Net \cite{R2AU-Net} for better segmentation accuracy. Our quantitative and qualitative results show that jointly training both tasks, using UNSB for image translation and R2AUNet for segmentation, positively influenced the output of the segmentation task. A major challenge in CT-based ventricle segmentation is the poor visibility of the target structure and soft tissue anatomies compared to those in T1w MRIs. The UNSB outperforms GAN-based methods in managing cross-domain mismatches, as demonstrated with natural images \cite{unpairedDiff} and confirmed by our findings. Due to training with unpaired images, our synthetic CTs have sharper details than the real ones, resulting in lower SSIMs when compared to synthetic and real MRIs in CT-to-MRI translation scenarios. To enhance segmentation, we adopted the R2AUNet model\cite{R2AU-Net}, incorporating recurrent and residual blocks, which boosted performance in both end-to-end and two-stage training settings. Comparing segmentation accuracy with uncertainty, the end-to-end UNSB model showed stronger correlations, likely due to its higher-quality MRI-to-CT translation and reduced discrepancy between the real and synthetic CT images for the segmentation network, in comparison to GAN-based methods. We acknowledge a few limitations. First, we used MRI manual segmentations from experts for training and results from multi-atlas label-fusion in both training and testing stages. Multi-atlas label-fusion, used for silver ground truths, has excelled in clinical studies and biomarker derivation for neurological conditions, despite its slower speed. While full manual segmentation would be desirable, the obtained silver ground truths provided good results based on visual inspection. Second, with the computational cost of diffusion models, we relied on a slice-by-slice processing approach. As diffusion models for high-resolution 3D images are desirable for clinical applications, we will explore computationally efficient methods to adapt our approach to 3D in the future. Patients requiring ventriculostomy may face ventricular enlargement and intracranial hemorrhage. Therefore, this study initially used healthy seniors over 70 years with naturally enlarged ventricles. We plan to include more clinical cases to enhance our DL model's capability, potentially aiding other CT-based ventricle segmentation applications.

\section{Conclusion}
We have proposed a novel, uncertainty-aware, diffusion-based CT brain ventricle segmentation method without the need for expert labels in the target image modality. Furthermore, we have demonstrated the benefits of diffusion Schrödinger Bridge and joint image translation and segmentation training against other approaches in a comprehensive comparison. As well as being beneficial to ventriculostomy surgical planning, we hope the proposed method would benefit other clinical applications that require ventricle assessment.

\noindent \textbf{Acknowledgment}. We acknowledge the support of the New Frontiers in Research Fund (NFRF) (fund number: NFRFE-2020-00241)

\noindent \textbf{Disclosure of Interests}
The authors declare no competing interests.

%
%
%
\bibliographystyle{splncs04}
%

\end{document}